# A Baseline Model for Software Effort Estimation

Peter A. Whigham, Caitlin A. Owen and Stephen G. MacDonell
*Department of Information Science, University of Otago,
PO Box 56, Dunedin 9054, New Zealand*
*{peter.whigham, stephen.macdonell}@otago.ac.nz*

**Abstract**

*Software effort estimation (SEE) is a core activity in all software processes and development lifecycles. A range of increasingly complex methods has been considered in the past 30 years for the prediction of effort, often with mixed and contradictory results. The comparative assessment of effort prediction methods has therefore become a common approach when considering how best to predict effort over a range of project types. Unfortunately, these assessments use a variety of sampling methods and error measurements, making comparison with other work difficult. This article proposes an automatically transformed linear model (ATLM) as a suitable baseline model for comparison against SEE methods. ATLM is simple yet performs well over a range of different project types. In addition, ATLM may be used with mixed numeric and categorical data and requires no parameter tuning. It is also deterministic, meaning that results obtained are amenable to replication. These and other arguments for using ATLM as a baseline model are presented, and a reference implementation described and made available. We suggest that ATLM should be used as a baseline of effort prediction quality for all future model comparisons in SEE.*

Categories and Subject Descriptors: D.2.9 [Software Engineering]: Management—*Cost estimation*; D.2.8 [Software Engineering]: Metrics—*Performance measures*; G.1.2 [Numerical Analysis]: Approximation—*Linear approximation*
General Terms: Algorithms, Performance
Additional Key Words and Phrases: Baseline model, transformed linear model

## 1. INTRODUCTION

Historically, the key objective of *software effort estimation* (SEE) has been to produce a robust, accurate predictive cost model [Albrecht and Gaffney 1983; Boehm 1981]. This is a challenging objective: the complex nature of software development has been recognized since the earliest efforts to characterize and predict software costs [Nelson 1966]. Unfortunately, the diversity of software projects, the inherent uncertainty associated with many of the explanatory variables, the interaction of intangible exogenous and endogenous mechanisms, and the nonstationary characteristics of the software development environment mean no single method will likely be the best for all project types [Collopy 2007; Menzies et al. 2006]. This has led to the ongoing publication of new, often complex, modelling approaches for effort estimation, increasingly based on approaches from the field of machine learning in light of their apparent ability to better cope with the complex circumstances just described. However, although these publications often compare the performance of the resulting models with other methods, the results are highly variable [Myrtveit and Stensrud 2012; Shepperd and Kadoda 2001], sometimes contradictory [Menzies and Shepperd 2012], and often difficult to interpret and validate [Kitchenham and Mendes 2009]. This is due in part to a number of factors: the use of datasets that are difficult to obtain or are proprietary and therefore not publicly available [Mair et al. 2005]; bias in dataset selection [Kitchenham and Mendes 2009]; initial data preparation methods that remove outlier examples and/or explanatory variables, making comparison with previous work difficult [Bosu and MacDonell 2013]; the sampling methods used when assessing model quality; the presented error measurements [Myrtveit and Stensrud 2012; Myrtveit et al. 2005]; bias in modeller expertise and parameter tuning [Song et al. 2013]; and a lack of meaningful comparative models and statistics. Although these issues arise in many fields where inductive model building is performed, and have been recognised as relevant to software effort estimation, they appear both common and enduring in this domain. Quality research in SEE requires these issues be managed and, where possible, addressed.

The purpose of this article is to develop a baseline model for SEE. Although *multiple linear regression* (MLR) has been used with mixed results in SEE and so has "fallen out of favour" in deference to other (machine learning) methods, we contend that with appropriate, automatic, data-driven transformations, MLR is an appropriate candidate for such a baseline. More than this, we argue (in the next section) it has particular characteristics that make it more amenable to a baseline role than many other methods. A brief historical examination of the use of MLR models in SEE is therefore appropriate to demonstrate some of the previously identified issues with model comparison and development. The reader is directed to Jorgensen and Shepperd [2007], Dejaeger et al. [2012], Minku and Yao [2013], and Ferrucci et al. [2014] for a general review of SEE and the use of machine learning methods in this context.

Early methods in effort prediction included the *constructive cost* (COCOMO) model [Boehm 1981] and *function point analysis* (FPA) [Albrecht and Gaffney 1983]. In formulating COCOMO, Boehm commented that "there are too many nonlinear interactions in software development for a linear model to work very well" [Boehm 1981, page 331]. This led to many model formulations that were multiplicative with associated power laws. Note, however, that log(a) + log(b) = log(a.b), and therefore logarithmic transformations between selected explanatory variables allow a multiplicative factor to be introduced to a linear model. There has also been some research suggesting that an exponent term may not increase model accuracy, although these results are somewhat contentious [Kitchenham 1992, 2002]. A review of early models may be found in Boehm [1981].

Since the work of Boehm [1981], studies have reached different conclusions as to which type of model produces the most accurate software effort estimations. For example, Dejaeger et al. [2012], Briand et al. [2000], Jeffery et al. [2000], and MacDonell and Shepperd [2003] found that MLR (with appropriate transformations) was the most accurate model in their empirical analyses. In contrast, Finnie et al. [1997] found that *artificial neural networks* (ANNs) and *case-based reasoning* (CBR) performed better than MLR. Mair et al. [2000], Tronto et al. [2006], and Park and Baek [2008] found that ANNs outperformed MLR, while Minku and Yao [2013] showed that a Pareto ensemble of ANNs had the best performance compared with a variety of methods. Heiat [2002] found that, sometimes, linear regression and ANNs exhibit similar performance but that this was not always the case. Analogy-based approaches [Chiu and Huang 2007] have also been shown to produce more accurate models than MLR, ANNs, simple decision trees, and stepwise regression [Shepperd and Schofield 1997] in other studies.

Why do these discrepancies exist? One obvious reason is that some datasets are more suited to some types of models than others, and therefore different datasets will produce different model performances. In addition, as argued by Kitchenham and Mendes [2009], naively applying a linear regression model is often inappropriate due to the non-normality of the residuals and without considering transformations to the response and/or explanatory variables. (More generally, the naive or uninformed application of any method may lead to misleading results and conclusions.) In particular, Kitchenham and Mendes [2009] argued that a comparison with linear regression "will be scientifically valueless, if the regression techniques are applied incorrectly." They went on to demonstrate that, with appropriate transformations and handling of categorical data, linear regression can produce suitable and accurate predictive models. Although some studies have used log-transformed variables (e.g., Finnie et al. [1997], Jeffery et al. [2000], Chen et al. [2005], and Dejaeger et al. [2012]), Jorgensen and Shepperd [2007] found that few studies look at the "properties of the data." Although the warnings from Kitchenham and Mendes [2009], were presented some years ago, it appears that the trend in applying linear models without a consideration of the data and model assumptions has continued [Bardsiri et al. 2014], or that no comparison against other models or the use of a training/test comparison is even considered [Aljahdali and Sheta 2013]. In response, in recent years increasing attention has been directed to the need for baseline models in SEE, against which any newly proposed model should be positively compared before adoption and use, that is, the new model should be shown superior to the baseline model. This article addresses the issue of baseline models for software effort estimation to provide a meaningful measure of model quality. This will address several of the previously noted issues by enabling a reproducible, comparative baseline assessment for any new model and data combination.

The remainder of this article is structured as follows: Section 2 introduces the characteristics of baseline models and describes the proposed baseline linear model for use in SEE. In addition, a baseline error measurement is introduced to complement the data-independent role of baseline concepts. Section 3 compares the proposed baseline model to two examples from the literature, showing that the model satisfies the characteristics for an appropriate baseline. Section 4 presents further arguments for requiring a baseline model in SEE and discusses the expected role of the baseline model. Section 5 concludes the article and summarises the contributions of this work. A reference implementation in the R programming language is provided in the Electronic Appendix to this article, which can be accessed via the ACM Digital library.

## 2. BASELINE MODELS

A baseline model defines a meaningful point of reference and, in order to be both useful and widely used in SEE, should possess a number of characteristics. In short, it should:

(1) be simple to describe, implement, and interpret;

(2) be deterministic in its outcomes;

(3) be applicable to mixed qualitative and quantitative data;

(4) offer some explanatory information regarding the prediction by representing generalised properties of the underlying data;

(5) have no parameters within the modelling process that require tuning;

(6) be publicly available via a reference implementation and associated environment for execution; and

(7) generally be more accurate than a random guess or an estimate based purely on the distribution of the response variable.

Any field of study where new methods for prediction are presented should require a comparison with a known baseline model. This would ensure that the predictive properties of a method can be adequately compared and assessed. In addition, the ongoing use of a baseline model in the literature would give a single point of comparison, allowing a meaningful assessment of any new method against previous work. Work in SEE involving regression error curves [Bi and Bennett 2003; Mittas and Angelis 2012] and model comparison frameworks [Mittas et al. 2015] have included a naive baseline model prediction using the mean or median of the data. In addition, a baseline model using the mean of a set of random samplings (with replacement) has also been proposed [Shepperd and MacDonell 2012]. This model has been incorporated into an error measurement

(defined as the standardised accuracy measure, SA). Work on cross- versus within-company cost estimation [Kitchenham et al.2007] has also recommended the use of a naive baseline and that regression analysis be the default model.

Given the preceding characteristics for a baseline model, a number of possibilities exist: a linear regression model, a simple decision tree, generalised linear models, extensions to linear and tree-based modelling, and so on. However, given our requirement for simplicity and availability, a simple linear regression as defined in the R programming language [R Development Core Team 2011] is proposed here. Other deterministic methods may be suitable, although for our purposes we only require one method that can be easily used and implemented. Although the automatic application of MLR does not deal with issues such as collinearity between explanatory variables, if the performance is generally better than a single-valued or random approach (mean, median, guessing), this also satisfies our baseline requirements. MLR with appropriate automatic transformations can also perform well without any parameter tuning (as shown in Section 3). Our baseline model is presented as an implementation in the R programming environment. Since R is a free, open-source environment with a standard core MLR implementation, this is one appropriate avenue for the baseline implementation. The issues of an appropriate, automatic method for data transformation and handling of qualitative data will be addressed in the next section.

## 2.1 Automatically Transformed Linear Baseline Model (ATLM)

The multiple linear regression form that we will consider is

$$y_i = \beta_0 + \beta_1 x_{1i} + \beta_2 x_{2i} + ..\beta_n x_{ni} + \varepsilon_i, \quad (1)$$

where $y_i$ is referred to as the *quantitative* response variable, the $x_i$ are explanatory variables, and the $\beta_i$ are determined using a least squares estimator [Neter et al. 1996]. The standard contrasts approach of dummy variables is used for each qualitative $x_i$, since this is the default method in R [R Development Core Team 2011] for handling categorical explanatory variables in linear models. Since SEE data is often skewed, transformations to the response and/or explanatory variables are often appropriate when forming linear models [Kitchenham and Mendes 2009]. The base implementation (referred to as ATLM) will assess the suitability of log and square-root transformations of the response and explanatory variables based on the underlying distribution of the data. The more complex consideration of model residuals, use of Box-Cox transformations, or stepwise regression [Venables and Ripley 2002] which would typically be examined when developing a linear model will not be considered. This is a deliberate decision since ATLM is not intended to be the "best" model, and these added complexities often involve parameter settings and therefore violate our baseline characteristics introduced in Section 2. The general approach is shown in Algorithm 1. Transformations are calculated for the response and explanatory variables using the training data, applied (as needed, based on skewness) to the training and test data, and a linear model constructed and used to predict the test data, with the final predictions inverted to produce the original scaling of the estimated response. These predictions can then be used to calculate the model error.

ALGORITHM 1: Linear Model Prediction
**Input:** formula, training, test
**Output:** Predictions for test data

| | |
|---|---|
| *transforms* | ← **calculate.transforms** (*training*) |
| *trans.training* | ← **apply.transforms** (*transforms,training*) |
| | # Apply transformations |
| *trans.test* | ← **apply.transforms** (*transforms,test*) |
| *lm.see* | ← **lm** (*formula,trans.training*) |
| | # Determine linear model |
| *predictions* | ← **predict** (*lm.see,newdata = trans.test*) |
| | # Predict response |
| **return** (**invert.predictions**(*transforms,predictions*)) |
| | # Return untransformed predictions |

The transformation step is calculated by comparing the skewness [Dimitriadou et al. 2011; Joanes and Gill 1998] for each variable (the *b1* skewness measure from Joanes and Gill [1998, page 184], is implemented) when a logarithm, square-root, and no transformation are applied, as shown in Algorithm 2. The transformation that results in the least skewed data for each variable is selected and used when constructing the linear model and predicting effort. Note that a variable that is categorical is ignored in this step. The appropriate inverse transform of the predictions is applied so that the model results can be meaningfully compared to the untransformed test data.

ALGORITHM 2: Calculate.Transforms
**Input:** data, transformfns ={none,log,sqrt},
invtransformfns ={none, exp, sqr}
**Output:** Transformation table

*transformationTable* ← {}
**For each** *variable* **in** *data*
# for the response and each explanatory variable
   **If** *variable* **is a factor**
   #do nothing for categorical factors
       *transfn* ← *none*
       *invfn* ← *none*
   **Else**
   # determine transformation
   *skewness* ← {}
     **For each** *fn* **in** *transformfns*
     # apply each transformation to each variable
       *skewdata* ← *fn* (*data*)
       # apply transformation function
       *skewness* ← *skewness* ∪ **skew** (*skewdata*)
       # record skewness
   *best* ← **min** (*skewness*)
   # Best function has lowest skewed data
   *transfn* ← *transformfns* (*best*)
   *invfn* ← *invtransformfns* (*best*)
  *transformationTable* ← *transformationTable* ∪
  {*variable, transfn, invfn*}
**return**(*transformationTable*)

## 2.2 A Baseline Accuracy Measurement

A number of measures have been used in the literature for comparing model performance in SEE, including: *mean magnitude of the relative error* (MMRE) [Shepperd and Kadoda 2001], PRED(x), a measure of accuracy where predictions are within x% of the measured value, the *standardized accuracy* SA [Shepperd and MacDonell 2012], which compares a prediction against a mean of a random sampling of the training response values, *logarithmic standard deviation* (LSD), and other measures that form a relative assessment of performance by dividing the prediction by the training/test response. However, apart from SA, none of these measures gives a baseline value that can indicate whether the predictions are better than a random or constant prediction. A standard measure that gives a value of 1 for a model that predicts a constant value (such as the mean or median of the response variable) is nonnegative, and for any useful predictor will be less than 1, and is defined as [Breiman 1993]

$$RE^* = var(residuals) / var(measured),$$

where *var(residuals)* is the variance of the residuals (i.e., predicted − measured), and *var(measured)* is the variance of the measured response. $RE^*$ is therefore an appropriate baseline error measurement since any model that produces an $RE^*$ greater than 1 would be considered poor, independent of the dataset.

## 3. BASELINE EXPERIMENTS

This section will give two examples using the baseline implementation ATLM. The intention is to show that ATLM satisfies the properties stated in Section 2. In terms of training and testing a model, there are a number of different approaches commonly used in SEE. Menzies et al. [2006] argue that a small test set size should be used because "that is how they will be assessed in practice." Some studies have used a test set size of 10, including Wittig and Finnie [1997], Mair et al. [2000, 2006], probably because this was used in an early work on effort prediction [Adrangi and Harrison 1987]. Other studies have used a test size of 1 (i.e., leave-one-out cross-validation), including Samson et al. [1997], Shepperd and Schofield [1997], Kitchenham et al. [2002], and, in particular, Myrtveit et al. [2005] who describe this as the "real-world situation." Recently there have been arguments presented for leave-one-out as the most appropriate validation method [Kocaguneli and Menzies 2013], although this recommendation has not yet influenced the field. Since there is no agreed standard method for assessing the quality of a model in SEE, the baseline implementation supports leave-one-out cross-validation, n-way cross-validation, and a sampling of any size for repeated training/test set measurements. Since we cannot anticipate the total range of error measurements of training/test regimes, the baseline model R code offers a simple interface taking a single training and test set. All runs of ATLM return the predicted and actual values as a table that can be exported for further analysis if required.

## 3.1 Ex 1: Randomized Training/Test Sets

The paper by Minku and Yao [2013] examined the concept of ensemble methods for software effort estimation. Their paper used a number of datasets and measured performance by randomly selecting, without replacement, 10 examples for testing with the remaining data used for training, as suggested by Menzies et al. [2006]. This was repeated 30 times to produce an overall estimate of accuracy. The error measurements MMRE, LSD, and PRED(25) were used to assess model quality. We calculate and report $RE^*$ for ATLM but, as we do not have the residuals for the ensemble method or ANN methods, we cannot calculate $RE^*$ for these models. To enable comparison we therefore also report LSD, MMRE, and PRED(25). ATLM was used to predict effort on three of the datasets analyzed by Minku and Yao [2013], that is, the Cocomo81, Desharnais, and OrgAll datasets, to demonstrate that the proposed baseline model produced predictions of comparable quality to those of the complex methods proposed in this article (Table II). The Cocomo81 dataset [Menzies et al. 2012] used 16 numerical features: 15 cost drivers and lines of code; in addition, the categorical development type (Mode) parameter was used as input. Following the default approach of "R" and that of Minku and Yao [2013], this categorical feature was handled using the approach of dummy variables [James et al. 2013]. The predicted value was total actual effort in person months. The Desharnais dataset was reduced following the data preparation steps of Minku and Yao [2013], resulting in the removal of four entries with missing values and three other entries which they had determined as outliers. In addition, the Language Type (Development Type) variable, coded as an integer value 1, 2, or 3 representing the language types Cobol, Advanced Cobol, and 4GL, was converted to a factor so that it was handled appropriately by ATLM. It is worth noting that the Non-Adjusted Points variable was included in the model even though it is a linear combination of the two variables Transactions and Entities. Although this is not good practice when constructing a linear model, this redundant feature was included so that a direct comparison using the same data and variables could be made. The model predicted actual effort in person hours.

Table I shows that simple ATLM is comparable with the Ensemble or ANN models from Minku and Yao [2013]. The greyed cells indicate better values, though not necessarily significantly better. No statistical comparison between the results will be considered since it is clear that ATLM performs with a similar (or better) mean and variance to those of the Ensemble/ANN method. In addition, for each dataset and each run of 30 cross-validation folds, different results were produced. Although these were not necessarily statistically different, the variation highlights the issue that we are discussing: that a fair model comparison can only occur when the same training/test splits are used. Note that obtaining an $RE^*$ value close to 1 (with high variance) for the OrgAll dataset suggests this dataset is very difficult to model. This issue, which lends further support for the use of a baseline model, will be examined in Section 4.

Table I. Test Performance Comparison between ATLM and Ensemble/ANN Method with Training/Test Split

| Data Set | Linear Model ATLM (mean ± std.dev.) | | | | Best Ensemble/ANN (mean ± std.dev.) | | |
|---|---|---|---|---|---|---|---|
| | LSD | MMRE | PRED(25) | RE* | LSD | MMRE | PRED(25) |
| Cocomo81 | 0.54 ± 0.15 | 0.44 ± 0.13 | 0.43 ± 0.18 | 0.30 ± 0.27 | 1.92 ± 0.77 | 2.79 ± 0.67 | 0.16 ± 0.13 |
| Desharnais | 0.47 ± 0.10 | 0.39 ± 0.12 | 0.44 ± 0.14 | 0.85 ± 0.96 | 1.03 ± 1.21 | 0.47 ± 0.19 | 0.42 ± 0.18 |
| OrgAll | 0.83 ± 0.26 | 0.73 ± 0.33 | 0.30 ± 0.15 | 0.87 ± 1.32 | 1.10 ± 0.95 | 0.73 ± 0.28 | 0.26 ± 0.13 |

Table II. Tenfold Cross-Validation Using ATLM (mean ± std.dev.)

| Dataset | Linear Model ATLM | | | | Hybrid ABE-PSO | |
|---|---|---|---|---|---|---|
| | LSD | MMRE | PRED(25) | RE* | MMRE | PRED |
| Cocomo81 | 0.54 ± 0.2 | 0.45 ± 0.26 | 0.41 ± 0.25 | 0.68 ± 1.1 | ∼0.5 | ∼0.4 |
| Maxwell | 0.58 ± 0.2 | 0.48 ± 0.17 | 0.37 ± 0.12 | 0.53 ± 0.8 | ∼0.6 | ∼0.37 |

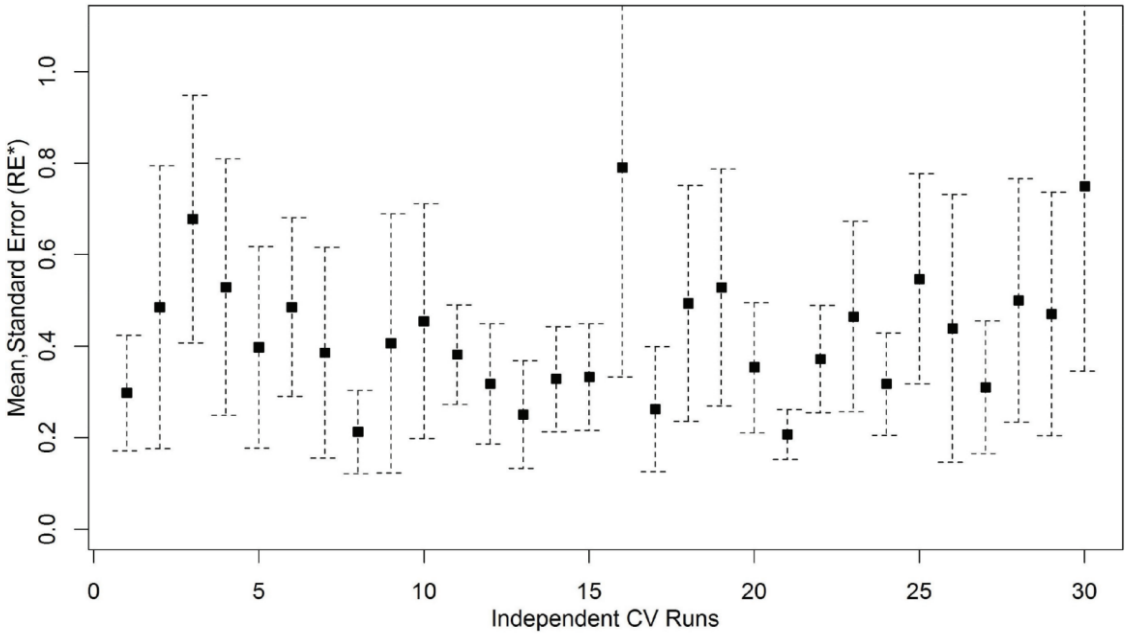

Fig. 1. Variation of RE* between runs of tenfold cross-validation using ATLM.

## 3.2 Ex 2: Cross-Validation

A recent paper by Bardsiri et al. [2014] proposed a hybrid method for SEE incorporating analogy-based estimation (ABE), clustering, and particle swarm optimisation (PSO). The authors concluded that the method could "considerably improve the accuracy of estimates" [Bardsiri et al. 2014, page 858]. A number of benchmark datasets were considered, including Cocomo81 and Maxwell. A (single) tenfold cross-validation assessment was used to estimate the errors associated with their proposed model.

Table II shows the results for a single tenfold cross-validation calculated as the mean and standard deviation over the errors for the 10 folds for ATLM and an estimate of the MMRE and PRED(25) results for the hybrid method of Bardsiri et al. [2014]. Unfortunately, no tables were given for the hybrid ABE-PSO error measures so these values have been estimated from their graphical (bar plot) figures. Note that the proposed baseline error measure RE∗ is also shown for ATLM. Of interest here is that the hybrid method was argued to be better than artificial neural networks, a standard *classification and regression tree* (CART) method, multiple linear regression, and stepwise regression. Our results would suggest that the ATLM is comparable with their hybrid method, even though the model is simple and has no parameters that require tuning from one dataset to another. This lends further support for the need to have a simple baseline model that can perform adequately over a range of datasets.

## 4. DISCUSSION

Our intent in developing ATLM is to ensure that new effort estimation methods are assessed in such a way that a fair comparison against a known standard can occur. In particular, there is a need for a baseline model that performs reasonably well on any dataset (unlike a prediction using the mean or median of the training data response variable) and that should be compared against the same training/test data folds as any proposed method. The issue of instability in software engineering datasets has been previously noted as an important aspect of modelling that must be considered [Turhan 2012]. Hence, unless a jackknife (leave-one-out cross-validation) method is used for assessing error, the variation in the sampling of the dataset may lead to misleading estimates of model quality. Unfortunately, a jackknife procedure tends to produce a large variance for datasets with outliers, and therefore may not be suitable for SEE datasets.

Consider Figure 1 which shows the mean and standard error for $RE*$ over 30 independent tenfold cross-validations for the Cocomo81 dataset using the deterministic linear model ATLM. The variation between runs is a result of the cross-validation sampling procedure and therefore suggests that a single cross-validation may not provide a representative example of model performance. Given that this type of variation can occur with a deterministic model, the issues are likely compounded when the model being measured is stochastic. This is a strong argument for using a deterministic baseline model that uses the same training/test data when a comparison of model performance is required. This would at least allow the following statement to be made: "under this particular set of training/test sets there is a significant difference between our model and that of ATLM." Comparison with other methods that also showed a significant difference would at least be able to argue some useful prediction performance.

An alternative solution to this variation in performance is to perform many runs of cross-validation or training/test splits (many more than 30) to obtain a fair estimate of mean/variance in performance and to allow a meaningful statistical comparison. Unfortunately, many of the complex methods considered for SEE are computationally expensive, and therefore this solution is often not feasible.

## 5. CONCLUSION

This article has argued that a baseline model is required for future developments in SEE when a new method or comparison of existing methods is undertaken to evaluate predictive quality. The proposed linear model with automatic transformations (ATLM) has been shown to satisfy the requirements for such a model, and should therefore be used as the baseline model when proposing and assessing new models of effort estimation.

## ELECTRONIC APPENDIX

The electronic appendix containing the R code and supporting documentation for this article can be accessed in the ACM Digital Library. These materials are also available through the PROMISE data repository.

## ACKNOWLEDGEMENTS

This work was partly supported by an Otago University School of Business Summer School grant.

The authors would like to thank Dr. Grant Dick for suggestions regarding the R implementation. Professor Martin Shepperd must also be thanked for constructive comments of a draft of this manuscript. The OrgAll dataset was kindly supplied by Dr. Leandro Minku.